\documentstyle[aps,epsfig,preprint]{revtex}
\author{Mauro Sellitto and Mario Nicodemi} 
\address{Dipartimento di Scienze Fisiche, Unit\`a INFM -- 
Universit\`a ``Federico II'' \\
 Mostra d'Oltremare, Pad. 19 -- 80125 Napoli -- Italy \\
E-mails: {\tt sellitto@na.infn.it} and {\tt nicodemim@na.infn.it} }
\author{Jeferson J. Arenzon} 
\address{Instituto de F\'{\i}sica -- UFRGS \\ CP 15051 -- 91501-970 --
Porto Alegre -- RS -- Brazil \\ E-mail: {\tt arenzon@if.ufrgs.br}}
\title{The Blume-Emery-Griffiths Spin Glass Model}
\date{\today}
    
\tighten

\newcommand{\tr}{\mathop{\mbox{Tr}}}
\newcommand{\Dz}{{\cal D}z\:}
\newcommand{\Dzo}{{\cal D}z_0\:}
\newcommand{\Dzu}{{\cal D}z_1\:}
\newcommand{\virgula}{\;\;,}
\newcommand{\ponto}{\;\;.}
\newcommand{\e}{\mbox{e}}

\begin{document}
\maketitle

\begin{abstract}
The equilibrium properties of the Blume-Emery-Griffiths model
with bilinear quenched disorder are studied for the case of 
attractive as well as repulsive biquadratic interactions.
The global phase diagram of the system is calculated in the context 
of the replica symmetric mean field approximation.
\end{abstract}   

\section{Introduction}
    
The Blume-Emery-Griffiths (BEG) \cite{BEG} model has been successfully 
applied 
to explain the behavior of different physical systems such as
He$^3$-He$^4$ mixtures, microemulsions, semiconductor alloys, 
to quote only a few.
In this paper we investigate the effect of quenched disorder on 
the mean field phase diagram of a version of the BEG
model in which orientational and particle degrees of freedom 
are explicitly introduced.
We consider the Hamiltonian:
\begin{equation}
{\cal H} =  - \sum_{i<j} J_{ij} S_i S_j n_i n_j 
            - \frac{K}{N}\sum_{i<j}  n_i n_j  
            - \mu \sum_i n_i  \virgula
\label{H}
\end{equation}                                  
where $S_i=\pm 1$, $n_i=0,1$; the bilinear couplings $J_{ij}$ are 
quenched Gaussian random variables with average $J_0/N$ and variance 
$J^2/N$, and the sign of biquadratic interaction may be negative.
This Hamiltonian represents a general framework to study the 
complex behavior of many physical systems. 
In the simple case of pure systems  it describes,
for example, a fluid model with magnetic properties like those of 
polar liquids, which exhibits multiple exotic phase diagrams  
\cite{Tanaka,Salinas89,HB,Branco,Akheyan}. 
In presence of quenched disorder the Hamiltonian (\ref{H}) implements
different models whose rich phase diagrams are not yet fully explored.
Some limiting cases include the standard Ising spin glass \cite{MPV}
($\mu\to\infty$; $n_i=1\; \forall i$) and the Ghatak-Sherrington model 
\cite{GS,Salinas94,LagedeAlmeida,MottishawSherrington,Yokota,Urahata}
($K=0$). In the particular case of $K=-1$ we 
recover the Frustrated Ising Lattice Gas \cite{Arenzon,NC} that is 
a version of the {\em Frustrated Percolation} model \cite{Coniglio},
whose properties suggest a possible close connection 
with the theories of {\em structural glasses} \cite{Kirk}.
Essentially, the model considers a lattice gas in a frustrated medium
where the particles have an internal degree of freedom (given by
its spin) that accounts, for example, for possible orientations of
complex molecules in glass forming liquids. These steric effects
are greatly responsible  for the geometric frustration appearing
in glass forming systems at low temperatures or high densities.
Besides that, the particles interact through a potential that may be 
attractive or repulsive depending on the value of $K$. 

Here we try to further 
elucidate the phase diagram properties of Hamiltonian 
(\ref{H}) within a replica mean field theory approach. 
Specifically, the presence of disorder in the bilinear term  
leads to the appearance of a transition from a
paramagnetic to a spin glass phase at low temperature or high
density. When  the biquadratic interaction is attractive
or weakly repulsive we found, accordingly with the value of 
density, two spin glass transitions
of different nature separated by a tricritical line: 
at high density the transition is continuous while at 
sufficiently low density the transition becomes discontinuous.
When the particle repulsion is rather strong
two new phases with a sub-lattice structure may appear:
an antiquadrupolar phase and at lower temperatures, an antiquadrupolar 
spin glass phase.
The antiquadrupolar spin glass transition line is always continuous  
whatever is the value of the chemical potential.
When the bond weights are not symmetrically distributed, 
a ferromagnetic phase (as well as a ferrimagnetic one, 
depending on the value of $J_0$) may appear in the phase diagram.
One can also identify a line where a dynamical instability appears,
as in spin glass models with discontinuous transition 
like the Potts glass \cite{softpotts,Felix} 
and the $p$-spin model \cite{Kirk,pspin}.
Moreover, preliminary results on the stability analysis indicates the 
presence of a transition line from a `phase' with one to a `phase' with 
infinity replica symmetry breaking.
This rich scenario and the unification picture provided by the model
are the main motivations of this paper.

\section{The phase diagram}

Many general properties of the phase diagram of Hamiltonian (\ref{H})
may be observed by simply studying the replica symmetric solutions of
the mean-field equations. In the following paragraphs
we discuss  how the many thermodynamic phases appear 
when the parameters of Hamiltonian~(\ref{H}) are changed.
 
For the sake of clarity we work out separately the cases of attractive, 
$K > 0$, and repulsive, $K < 0$, biquadratic interaction.

\subsection{Attractive biquadratic interaction}
  				  
Using the replica method the free energy can be computed as:
\begin{eqnarray}
\beta f &=& - \lim_{n \rightarrow 0} \frac{1}{n} \ln [Z^n]_{av}  \virgula
\end{eqnarray}                                    
where $n$ is the numbers of replicas and $[\cdots]_{av}$ denotes the
average over the disorder. We obtain:
\begin{eqnarray}
\beta f &=& \lim_{n \rightarrow 0} \frac{1}{n} \left[
  \frac{1}{2} \beta^2 J^2 \sum_{a<b} q_{ab}^2 
  + \frac{1}{2} \beta J_0 \sum_a m_a^2   \right. \nonumber \\
  && \left. + \frac{1}{4} (\beta^2 J^2  
   + 2 \beta K) \sum_a d_a^2  - \ln Z' \right] \virgula
\end{eqnarray}
where $Z'$ is the single site partition function:
\begin{eqnarray} 
Z' &=& \tr_{S,n}  \exp\left\{
   \beta^2 J^2 \sum_{a<b} q_{ab} S^a n^a S^b n^b   
   + \beta \sum_a J_0 m_a  S^a n^a  \right. \nonumber  \\
&&  \left. + \sum_a \left[ \left(\frac{\beta^2 J^2}{2} + \beta K\right) d_a 
    + \beta \mu \right] n^a  \right\}  \virgula
\end{eqnarray}
and the order parameters: density ($d_a$), magnetization ($m_a$), 
and Parisi overlap ($q_{ab}$), are defined as
\begin{eqnarray} 
d_a&=& \left \langle n^a   \right \rangle \virgula \\
m_a&=&  \left \langle S^a n^a   \right \rangle    \virgula \\
q_{ab}&=&  \left \langle S^a n^a S^b n^b  \right \rangle   \ponto  
\end{eqnarray}
In the replica symmetric approximation,  
$m_a=m$, $d_a=d$ and $q_{ab}=q(1-\delta_{ab})$, 
one finds: 
\begin{eqnarray}
\beta f &=& \frac{1}{2} \beta J_0 m^2
            -\frac{1}{4} \beta^2 J^2 q^2
            + \frac{1}{4} (\beta^2 J^2 + 2 \beta K) d^2 \nonumber \\
        &-& \int \Dz \ln 2 \left[1 + \e^{\Xi} 
            \cosh (\beta J z \sqrt{q} + \beta J_0 m) \right]  \virgula
\end{eqnarray}                                    
where $\Dz={\rm d}z \e^{-z^2/2}$ is the standard Gaussian measure
and we define:
\begin{equation}
\Xi \equiv \frac{\beta^2 J^2}{2}  
(d-q) + \beta K d + \beta \mu
\ponto
\label{theta}
\end{equation}
The order parameters satisfy the saddle point equations:
\begin{eqnarray}
d&=& \int\Dz\frac{\cosh(\beta J z \sqrt{q} + \beta J_0m)}{\e^{-\Xi}  
  + \cosh(\beta J z \sqrt{q}+\beta J_0m)}  \virgula \\
q &=& \int\Dz   \frac{\sinh^2(\beta J z \sqrt{q} + \beta J_0m)}{
   [\e^{-\Xi} + \cosh(\beta J z \sqrt{q}+ \beta J_0m)]^2}  \virgula \\
m &=& \int\Dz\frac{\sinh(\beta Jz \sqrt{q} +\beta J_0 m )}{\e^{-\Xi}  
    + \cosh(\beta J z \sqrt{q} + \beta J_0m)}   \ponto
\end{eqnarray}
Notice that with a suitable transformation of our parameters 
($\beta\mu\to\beta\mu - \ln 2$) we recover the equations obtained
using spin-1 variables.

\subsubsection{The case $J_0=0$}
    
\begin{center}
\begin{figure}[h]
\epsfig{file=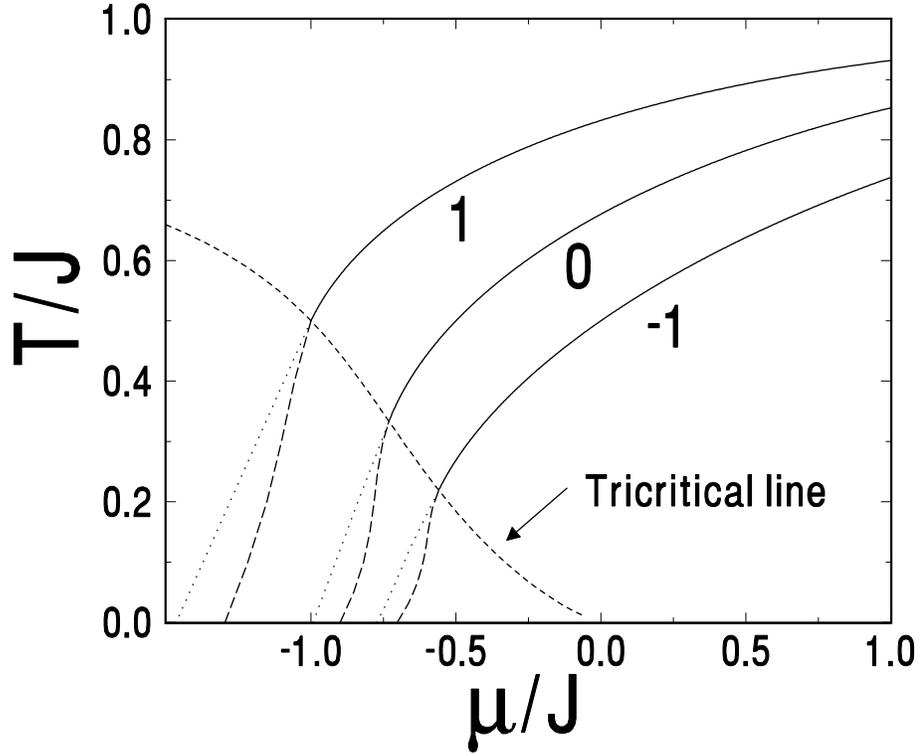,width=8.5cm,angle=270} 
\vspace{3.5cm}
\caption{Phase diagram for several values of $K/J$ showing the
paramagnetic (above) and spin glass phase (below the line). The 
continuous line
stands for the second order transition while the long dash line is the
discontinuous transition. Also depicted (dotted line) 
the line where the solution $q\neq 0$  first appears;
while the small dash line is a line of tricritical points.}
\label{pdmanyK}
\end{figure}
\end{center}

The simplest case in this situation corresponds to 
a zero disordered average, $J_0=0$. 
As can be seen in Fig.~\ref{pdmanyK}, where the phase diagrams
in the plane $(T,\mu)$ are depicted for several $K$, at
low temperatures or high values of the chemical potential
(high densities), there is a phase transition from a paramagnetic ($P$;
$m=q=0$) to a spin glass ($SG$; $m=0,\; q\neq 0$) phase. For positive values 
of $K$ (and $J_0=0$), these
are the only phases allowed to the system. They are separated 
by a continuous transition up to a tricritical point, $T_{trc}$, below 
which a discontinuous transition is observed. 
We just note that the Sherrington-Kirkpatrick (SK) model with its 
critical temperature
$T_c/J=1$ is recovered in two limits, both giving the highest possible
density ($d=1$): $\mu\to\infty$ or $K\to\infty$.

We locate the continuous transition by expanding the saddle point 
equations for small values of $q$, what leads to
\begin{equation}
\frac{J}{T_c}= 1+\exp\left(-\frac{K}{J}-\frac{J}{2T_c} 
-\frac{\mu}{T_c}\right) \ponto
\label{continuoustc}
\end{equation}
This equation is valid up to the tricritical point that may be located 
by expanding the single site partition function around the transition 
line where $q_{ab}=0$ and $d_a=\tilde d$ \cite{thankstoTheo}.
This leads to:
\begin{equation}
\frac{T_{trc}}{J}=\frac{-3+2K/J+\sqrt{9+4(K/J)^2-4K/J}}{4K/J} \ponto
\end{equation} 
For instance, if $K/J=-1$ we get $T_{trc}/J=(5-\sqrt{17})/4\simeq 0.219$ 
and $\mu_{trc}/J\simeq-0.559$\cite{Arenzon}. The line $T_{trc}(K)$
is also shown in Fig.~\ref{pdmanyK}.

\begin{center}
\begin{figure}[h]
\epsfig{file=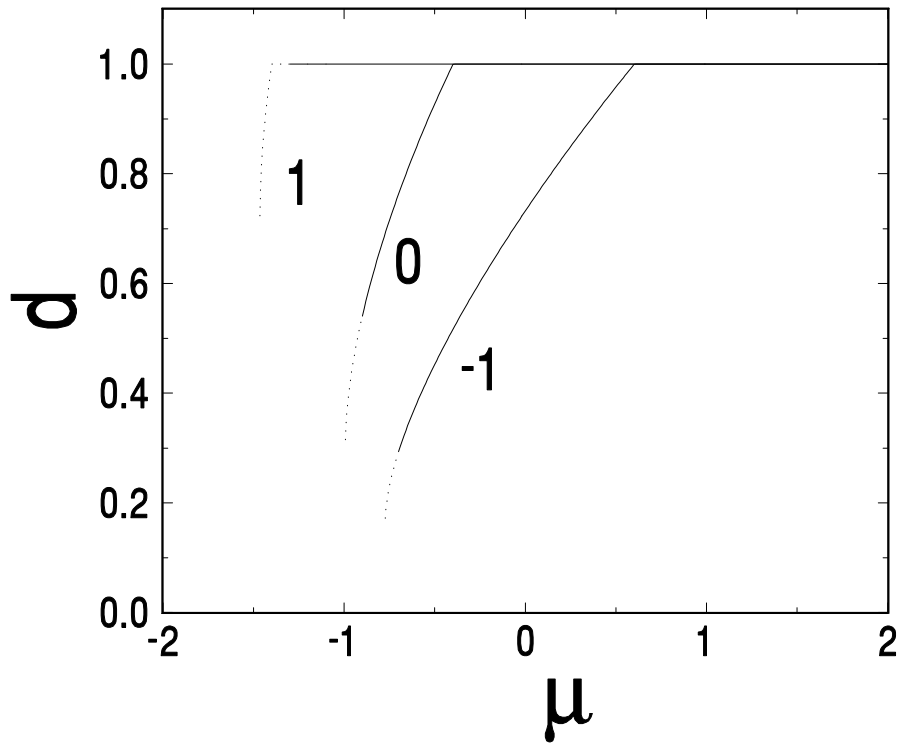,width=8.5cm,angle=270} 
\vspace{3.5cm}
\caption{The order parameter $d$ at $T=0$ for $K/J=-1,0$ and 1. 
The dotted part signals the region where non zero solution 
of saddle point equations first appears.
Observe that there is a particular value of the chemical potential, 
$\mu^*$, where the density becomes lower than 1.}
\label{dT0}
\end{figure}
\end{center}

The exploration of our equations in the $T=0$ limit offers a
simple tool to understand the subtle properties of the phase diagrams 
reported above. Moreover, remembering that our model shares
features with the frustrated percolation where no frustrated loops may be
closed, it is interesting to study the behavior of the density $d$ at $T=0$. 
This behavior is depicted in Fig.~\ref{dT0}. At $T=0$ the saddle point 
equations become:
\begin{equation}
d = \mbox{erfc} \left[ -\frac{1}{\sqrt{2d} } \left( 
    \frac{1}{2} C+ \frac{K}{J} d +\frac{\mu}{J} \right) \right]  \virgula
\end{equation}
where $C\equiv \beta J (d-q)$ and:
\begin{equation}
C = \sqrt{\frac{2}{d\pi}} \exp \left[-\frac{1}{2d} \left( \frac{1}{2} C +
    \frac{K}{J} d + \frac{\mu}{J} \right)^2 \right]    \ponto
\end{equation}
Actually, $d$ satisfies these equations up to $\mu^*$ given by 
\begin{equation}
\mu^* = -K -\frac{J}{\sqrt{2\pi}}    \virgula
\label{mustar}
\end{equation}
below which the close packing configuration is never achieved ($d<1$);
above $\mu^*$, $d=1$ and $C=\sqrt{2/\pi}$. 
The point $\mu^*$ (see Fig.~\ref{dT0}) individuates a clear cusp in the
behavior of $d(\mu)|_{T=0}$ and seems to be a characteristic value
for the system. 
We also observe that for high values of $K$, $K>\sqrt{2/\pi}$,
the region with $0<d<1$ disappears. 
    
From the tricritical point other two lines depart in the plane
$(\mu,T)$ (see Fig.\ref{pdmanyK}). 
They correspond to a thermodynamic 
line of discontinuous transitions from a paramagnetic to a SG phase, 
located where the free energy of the paramagnetic and spin glass
solution  are equal;
and a purely dynamical transition line whose presence 
is due to the large number of TAP metastable states \cite{TAP} 
which can trap the 
system for very long times (infinite time in the mean field model)
\cite{Heiko,KD}. 
The study of  this metastable glassy states within 
a suitable generalization of the TAP approach to the present model  
is in progress \cite{Maurothesis}.

In order to identify the dynamical transition line we should
compute the smallest eigenvalue (replicon) of the 1RSB stability 
matrix and impose the marginality condition \cite{marginal}.
An alternative procedure amounts to expand
the 1RSB free energy (see appendix) around the point
$m=1$, where $m$ denotes here the size of diagonal blocks in the 
Parisi ansatz \cite{Felix}. 
However at level of the approximation we work (replica symmetry) 
we can guess that the dynamical transition line is 
located near the line where 
a non zero solution of replica symmetric saddle point equations 
first appears; however, to check this point a more detailed 
investigation will be need.

When the temperature and chemical potential are close enough to
the discontinuous transition line we expect that the glassy phase of the 
system is exactly described by only one step of replica symmetry breaking
\cite{Maurothesis}. 
On the other hand, when $\mu\to\infty$ (the SK model), the full RSB scheme, 
with infinite steps, is needed in order to obtain the correct answer. 
Thus, a further `replica transition line' from 
a SG phase with one (or finite) replica symmetry breaking level 
to a SG phase with infinite levels of replica breaking is expected
to start at the tricritical point and terminate on the $T=0$ axis 
in some characteristic point, probably near $\mu^*$.
This transition line is expected to separate  qualitatively different
{\em aging} behaviors of the system \cite{CuKu,FraMe,CuDou}.

We can also apply a magnetic field to the system by including a
term like $-h\sum_i S_i n_i$ to the Hamiltonian eq.(\ref{H})
(this is done by adding the term $\beta m$ to the argument
of the hyperbolic functions in the saddle point equations).
The continuous transition is destroyed as soon as $h\neq 0$
and the first order transition ends in a critical point, the
system presenting the characteristic wing shaped phase diagram. 
The effects of a magnetic field, as well as the stability of
the RS solution in its presence were extensively studied for
the Ghatak-Sherrington model \cite{Urahata}.

\subsubsection{The case $J_0 \neq 0$}
    
\begin{center}
\begin{figure}[h]
\epsfig{file=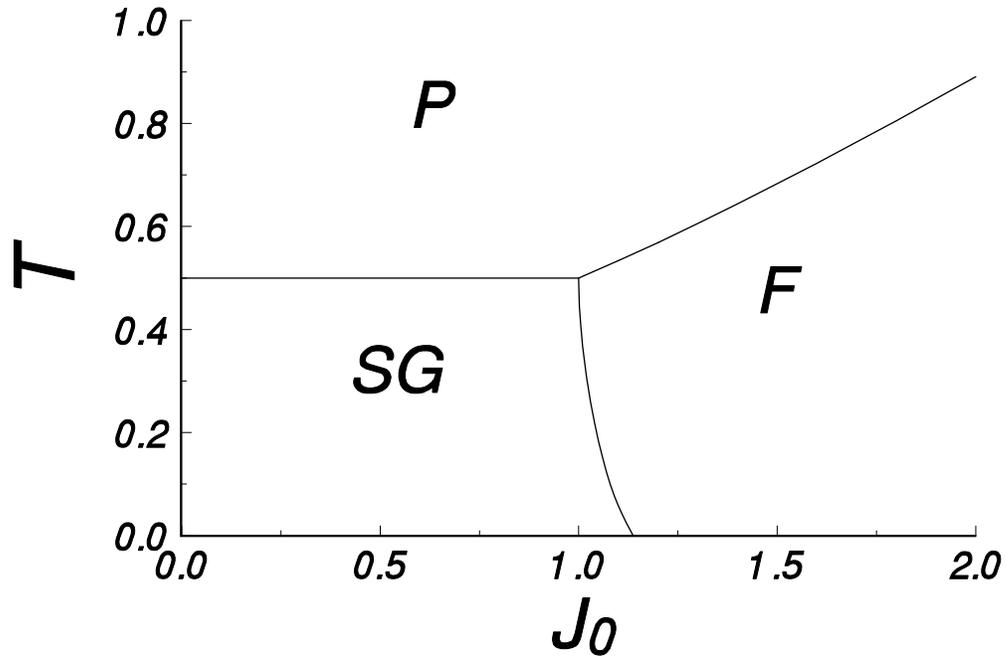,width=8.5cm,angle=270} 
\vspace{3.5cm}
\caption{Phase diagram for $K=-1$ and $\mu=0$.
The reentrance is an artifact of the RS approximation.
All transition lines are continuous and look much the same as 
the SK boundaries.}
\label{j0km1m0}
\end{figure}
\end{center}

Having given some details for the case $J_0=0$, we face below the more
general situation when $J_0$ is different from zero \cite{Schreiber}. 
There are now three kind of phases (see Fig.\ref{j0km1m0}): a paramagnetic
and a spin glass phase with
essentially the same characteristics we have discussed above, along with a 
ferromagnetic phase ($F$; $q\neq 0$, $m\neq 0$) which appears, as usual, 
in the low temperature region (or high $\mu$) when $J_0 \gg J$. 
As in the SK model, this ferromagnetic phase is reentrant (in RS), 
although here the degree of reentrance may vary.
    
When the transitions between these phases are continuous 
({\em e.g.}, Fig.\ref{j0km1m0}), the boundary lines may be analytically
obtained. The boundary between $F$ and $P$ is given by
\begin{equation}
\frac{J_0}{T_c} = 1 + \exp\left( -\frac{\mu}{T_c} -\frac{K}{J_0} -
\frac{1}{2J_0T_c}\right)     \virgula
\label{PF}
\end{equation}
while the boundary between $SG$ and $F$ obeys
\begin{equation}
T_c = J_0 (d-q)   \virgula
\end{equation}
where $d$ and $q$ are obtained solving the saddle point equations
with $m=0$. And the boundary between $SG$ and $P$ is our former result
eq.(\ref{continuoustc}).
Notice that in the limit $\mu$ (or $K$)$\to\infty$ , we recover the
SK boundaries ($T_c=J$ for $SG$-$P$, $T_c=J_0$ for $F$-$P$ and
$T_c=J_0(1-q)$ for $SG$-$F$).

\begin{center}
\begin{figure}[h]
\epsfig{file=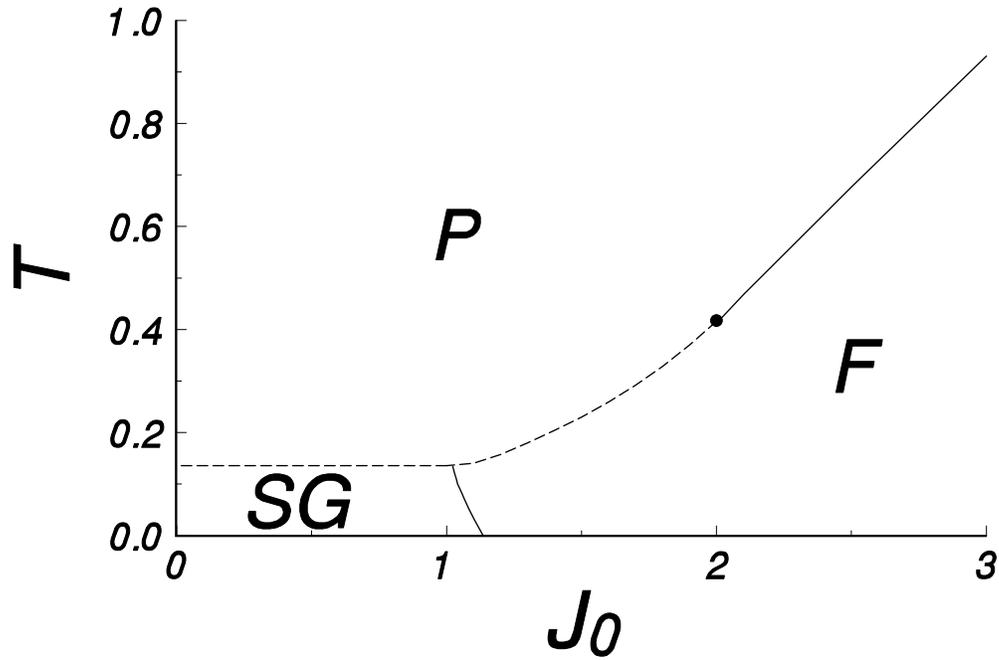,width=8.5cm,angle=270} 
\vspace{3.5cm}
\caption{Phase diagram for $K=-1$ and $\mu=-0.6$. 
The continuous lines are continuous
transitions while the dashed lines stand for discontinuous
transitions. Notice that the line $SG$-$F$ ends in an {\em endpoint}
and the transition $F$-$P$ is discontinuous up to the tricritical
point.}
\label{j0km1m06}
\end{figure}
\end{center}

Changing the value of the chemical potential the transitions may
become discontinuous, as can be seen in Fig.~\ref{j0km1m06} for
$\mu=-0.6$ and $K=-1$, where a tricritical point and an endpoint
show up.

As usual the analysis at $T=0$ is interesting. In this limit 
we get the following equations:
\begin{eqnarray}
d&=&\frac{1}{2}\mbox{erfc} \left( \frac{-\frac{1}{2} C-Kd-\mu-
    J_0 m}{\sqrt{2d}} \right) \nonumber \\ &&  + \frac{1}{2}
    \mbox{erfc} \left( \frac{-\frac{1}{2} C-Kd-\mu+J_0m}{\sqrt{2d}}
\right) 	\virgula  \\
m&=&\frac{1}{2}\mbox{erfc} \left( \frac{-\frac{1}{2} C-Kd-\mu-
    J_0 m}{\sqrt{2d}} \right) \nonumber \\ &&  - \frac{1}{2}
    \mbox{erfc} \left( \frac{-\frac{1}{2} C-Kd-\mu+J_0m}{\sqrt{2d}} 
		\right) \virgula
\end{eqnarray}
and
\begin{eqnarray}
C&=&\sqrt{\frac{2}{d\pi}} \exp\left[ -\frac{\left(\frac{1}{2}C+Kd+
   \mu\right)^2+J_0^2 m^2}{2d}\right] \nonumber \\
   && \cosh\left[- \frac{J_0m}{d}\left( \frac{C}{2}+Kd+\frac{\mu}{J} 
   \right)\right]  \virgula
\end{eqnarray}
    
We can then individuate the location of the continuous transition line, 
in the plane $(J_0,\mu)$, between the $F$ and $SG$ phases. It occurs 
where $m\to 0$ so, expanding the above equation in $m$, we get:
\begin{equation}
J_0 = \sqrt{\frac{\pi d}{2}} \exp\left[ \frac{\left(\frac{1}{2} C+Kd
      +\mu\right)^2}{2d} \right]    \ponto
\end{equation}
Here $d$ and $C$ are evaluated in the points where $m=0$. 
This equation is valid up to a given $\mu^*=-K-1/\sqrt{2\pi}$ (note
that since $m=0$ on this transition line, we just recover our former
result) and above this value, we find $J_0=\sqrt{\pi/2}\simeq 1.253$.
    
As is well known in the SK model, when $J_0$ grows too much, the SG phase
disappears. A similar phenomenon happens here, but the order of the
transition between the phases $F$ and $P$ depends on the value
of $\mu$, as expected. Notice that we are considering $K=-1$, but in
the next section we will show that for this value there are no 
antiquadrupolar (glassy or not)
phases and the characteristics presented here are also general for
greater values of $K$. We expect that for negative values of $K$ or 
$J_0$, there is also the possibility of having ferr{\em i}magnetic
or anti-ferr{\em o}magnetic ordering, respectively. The former is
characterized by having different magnetizations in the sublattices
($m_A\neq m_B$) and the latter by having opposite magnetizations 
($m_A=-m_B$). The analysis of these phases is beyond the scope of 
this paper.
    
\subsection{Repulsive biquadratic interaction}

\begin{center}
\begin{figure}
\epsfig{file=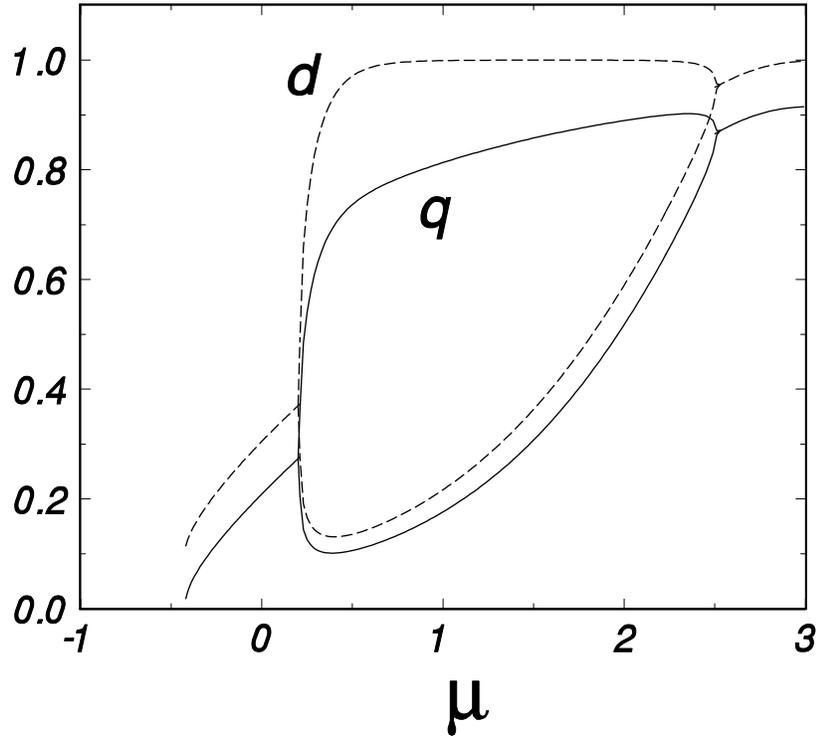,width=8.5cm,angle=270} 
\vspace{3.5cm}
\caption{Plot of the order parameters $q$ (solid line) and $d$ 
(dashed line) as a function of the
chemical potential $\mu$ for $K/J=-3$ and a fixed temperature
($T=0.1$). Both $SG$ and $AG$ phases are presented. In particular,
the AG solution is the region were both $q$ and $d$ are twofolded.
This figure is an horizontal cross section of the fig.
\protect{\ref{pdkm3}}.}
\label{qdsotto}
\end{figure}
\end{center}

Because the sign of the biquadratic coupling is now negative we 
have to take into account the possibility of further phase ordering 
within a two sub-lattice structure. This for instance is the case 
studied in refs.~\cite{Tanaka,Salinas89,HB,Branco,Akheyan} 
in absence of quenched disorder and in ref.\cite{daCosta} but
only for $\mu=0$. 
When $\mu\to\infty$ we recover the anti-ferromagnetic 
Sherrington-Kirkpatrick model, studied in ref.\cite{FKS}, and following
their prescription, the Hamiltonian can be rewritten as:
\begin{eqnarray}
{\cal H} &=& - \sum_{i,j} J_{ij} S_i^A S_j^B n_i^A n_j^B 
            - \frac{K}{N} \sum_{i,j} n_i^A n_j^B  \nonumber \\
         &&   -\mu \sum_{\alpha=A,B} \sum_i n_i^{\alpha} \virgula
\end{eqnarray}
where $\alpha=A,B$ is the index of the two sub-lattices.
Using the replica method we obtain the following free energy:
\begin{eqnarray}
\beta f &=& \lim_{n \rightarrow 0} \frac{1}{n} \left[
  \frac{1}{2} \beta^2 J^2 \sum_{a<b} q^{ab}_A q^{ab}_B 
  + \frac{1}{2} \beta J_0 \sum_a m^a_A m^a_B   \right. \nonumber \\
  && \left. + \frac{1}{4} (\beta^2 J^2  
   + 2 \beta K) \sum_a d^a_A d^a_B  
   - \frac{1}{2} \ln Z_A Z_B \right]  \virgula
\end{eqnarray}
where $Z_{\alpha}$ ($\alpha=A,B$) is the single site partition 
function of the sub-lattice $\alpha$:
\begin{eqnarray} 
Z_{\alpha} &=& \tr_{S,n}  \exp\left\{
   \beta^2 J^2 \sum_{a<b} q^{ab} S^a n^a
    S^b n^b + \beta \sum_a J_0 m^a  S^a n^a
   \right. \nonumber  \\
& &  \left. + \sum_a \left[ \left(\frac{\beta^2 J^2}{2} + \beta K\right) 
d^a + \beta \mu \right] n^a  \right\}   \ponto
\end{eqnarray}
In the replica symmetric approximation the free energy can be easily computed:
\begin{eqnarray*}
&&\beta f = \frac{1}{2} \beta J_0 m_A m_B
          -\frac{1}{4} \beta^2 J^2 q_A q_B
          + \frac{1}{4} (\beta^2 J^2 + 2 \beta K) d_A d_B \nonumber \\
     &&-  \frac{1}{2} \sum_{\alpha=A,B} \int \Dz \ln 2 
          \left[1 + \e^{{\Xi}_{\alpha}} 
          \cosh (\beta J z \sqrt{q_{\alpha}} + \beta J_0 m_{\alpha} ) 
	\right] \virgula
\end{eqnarray*}                                    
where 
\begin{equation}
\Xi_{\alpha} \equiv \frac{\beta^2 J^2}{2} 
(d_{\alpha}-q_{\alpha}) + \beta K d_{\alpha} + \beta \mu
\ponto
\label{thetaA}
\end{equation}
From the free energy, one derives the replica symmetric saddle point 
equations:
\begin{eqnarray}
d_A &=& \int\Dz\frac{\cosh(\beta J z \sqrt{q_B} + \beta J_0m_B)}{\e^{-{\Xi}_B}
    + \cosh(\beta J z \sqrt{q_B}+\beta J_0m_B)} \virgula \\
q_A &=& \int\Dz   \frac{\sinh^2(\beta J z \sqrt{q_B} + \beta J_0m_B)}{
    [\e^{-{\Xi}_B} + \cosh(\beta J z \sqrt{q_B}+ \beta J_0m_B )]^2} \virgula \\
m_A &=& \int\Dz\frac{\sinh(\beta Jz \sqrt{q_B} +\beta J_0 m_B )}{\e^{-{\Xi}_B}
    + \cosh(\beta J z \sqrt{q_B} + \beta J_0m_B)}  \ponto
\end{eqnarray}
The analogous equations for $d_B$, $q_B$ and $m_B$ are obtained from the 
previous one exchanging $A \leftrightarrow B$. A representative case
is shown in Fig.~\ref{qdsotto} where both $q_{A(B)}$ and $d_{A(B)}$
are presented in the $SG$ and $AG$ phases (see definitions below).
 
As stated above, new different phases emerge here.
At high temperatures (or low $\mu$), when the orientational degrees of 
freedom are not interacting ($q_A=q_B=0$), there are two possible
orderings: a paramagnetic (or quadrupolar) phase, $P$ ,with $d_A=d_B\neq 0$
and an antiquadrupolar one, $AQ$, with $d_A \neq d_B$.
The name quadrupolar is reminiscent from the spin-1 representation 
($\tau_i=S_i n_i$) where it labels the ordering of variables $\tau_i^2=n_i$. 
At low temperatures (or high $\mu$), the orientational
degrees of freedom become important and different glassy
phases show up. First, there is a spin glass phase ($SG$) with $d_A = d_B$ and
$q_A = q_B \neq 0$). This is the only glassy phase if the effects of particle 
repulsion are not very strong ($K$ not too negative). On the other side, 
when $K$ is highly negative, besides the $SG$ phase, there is also
an antiquadrupolar glass phase, $AG$, where the sub-lattice symmetry
is broken ($d_A \neq d_B$, $q_A \neq q_B$).

As in the previous section, when $J_0$ is allowed to be nonzero, 
the system develops a magnetization and 
additional, ordered phases may appear: when $J_0 \gg J$ at low temperatures
(or high $\mu$) a ferromagnetic phase ($F$; $m_A=m_B\neq 0$,
$d_A=d_B$, $q_A=q_B \neq0$) is encountered if $K$ is not too negative, 
or either a ferrimagnetic phase is entered ($I$; $m_A\neq m_B$, $d_A\neq
d_B$, $q_A \neq q_B$). Also, with $J_0\ll -J$, an anti-ferromagnetic
phase ($m_A=-m_B\neq 0$) may shows up.

\begin{center}
\begin{figure}
\epsfig{file=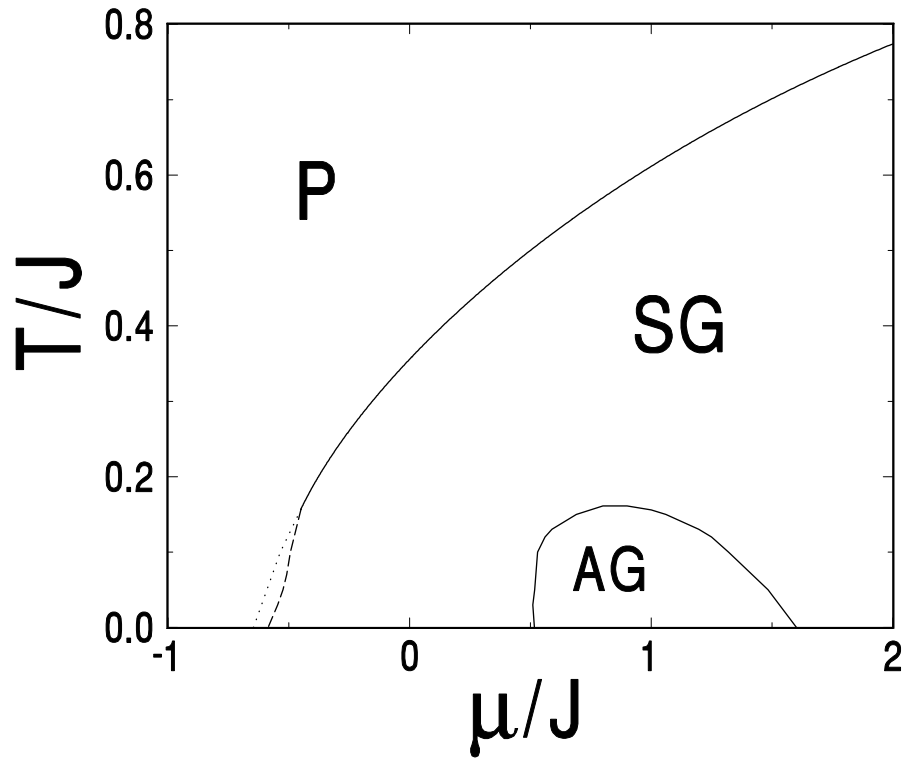,width=8.5cm,angle=270} 
\vspace{3.5cm}
\caption{Phase diagram for $K=-2$ showing the phase $AG$. Notice that
at this point there is no $AQ$ phase yet.}
\label{pdkm2}
\end{figure}
\end{center}

\begin{center}
\begin{figure}
\epsfig{file=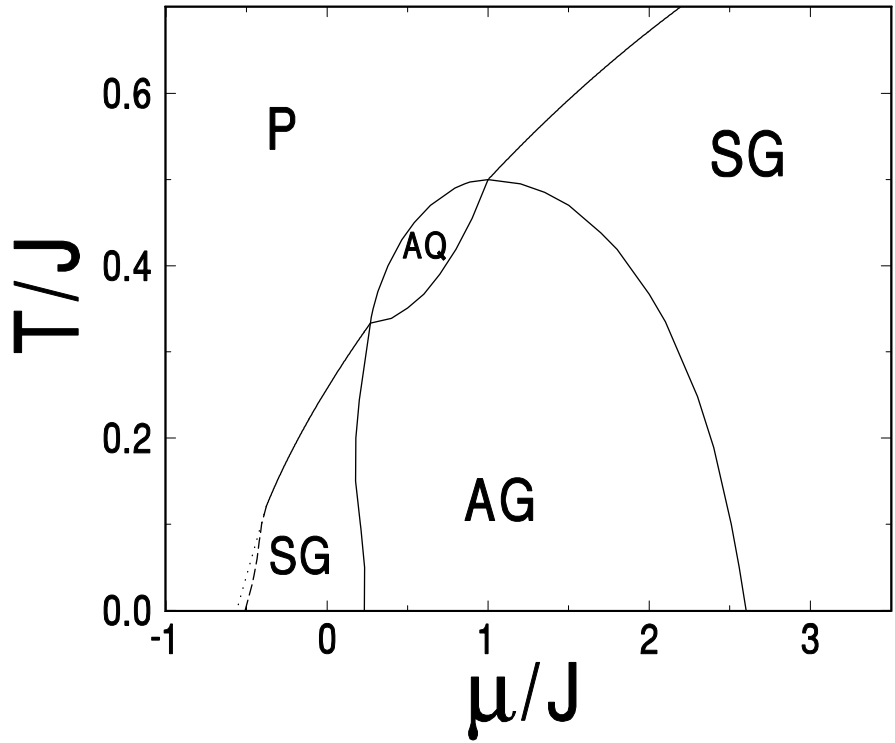,width=8.5cm,angle=270} 
\vspace{3.5cm}
\caption{Phase diagram for $K=-3$ where the $AQ$ phase can be seen.
This phase appears when the $AG$ border touches the $SG$-$P$
transition line for the first time.}
\label{pdkm3}
\end{figure}
\end{center}

\begin{center}
\begin{figure}
\epsfig{file=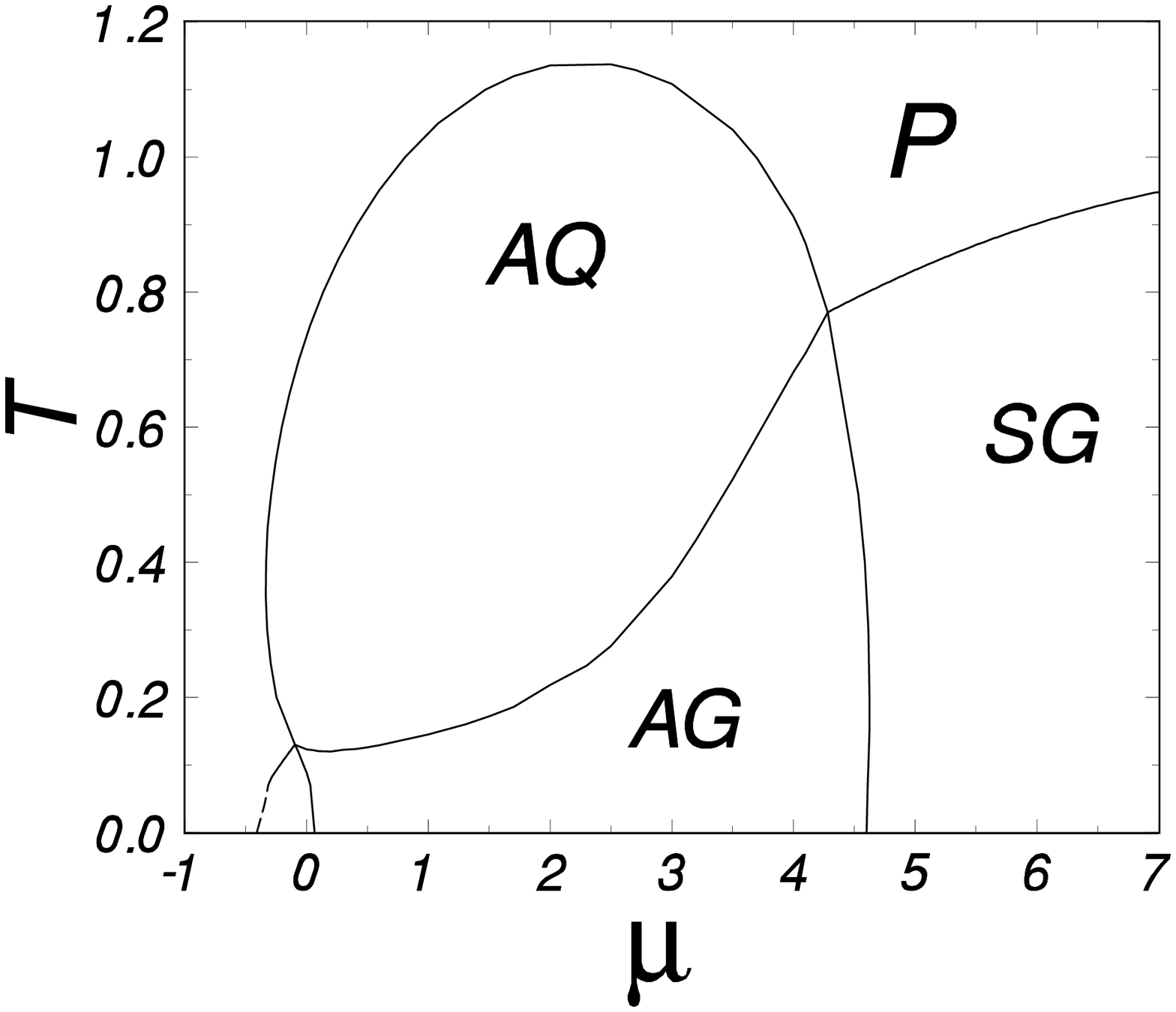,width=8.5cm,angle=270} 
\vspace{3.5cm}
\caption{Phase diagram for $K=-5$ showing the behavior of the various
phases for increasing values of $K$ (compares with figures
\protect{\ref{pdkm2}} and \protect{\ref{pdkm3}}).}
\label{pdkm5}
\end{figure}
\end{center}

From now on we consider $J_0=0$. As we decrease $K$ from a positive
value, there is a point, $K_{AG}\simeq -1.46$, where the $AG$ phase 
first appears, growing inside the previously described $SG$ phase
(see Fig.~\ref{pdkm2}). 
At a characteristic value, $K_{AQ} =-3/2-\sqrt{2}\simeq -2.91$,
an antiquadrupolar phase appears between the paramagnetic, the $SG$
and the $AG$ phases, as depicted in Fig.~\ref{pdkm3}. 
This value is obtained noticing that there are two points where
the $P$-$AQ$ line intercepts the $AQ$-$SG$ line (see
Fig.~\ref{pdkm3}), given by
\begin{equation}
T_{\pm}=\frac{1}{2K}\left( K+\frac{1}{2}\pm\sqrt{K^2+3K+\frac{1}{4}} 
\right) \ponto
\end{equation}
When both are equal, we get $K_{AQ}$. As soon as the $AQ$ phase appears,
the $SG$ phase is divided in two regions. At even lower values of $K$
(see Fig.~\ref{pdkm5}), 
the $AQ$ grows, while the left $SG$ phase shrinks and the right
one moves to higher values of $\mu$. This $SG$ phase disappears as
$K\to -\infty$. We can also notice that the $AG$ phase is invaded
by the $AQ$ as we approach this limit, and eventually we have
only the $P$ and $AQ$ phases, as we could expect since in this
limit we have a lattice gas with repulsive interactions.

With the same techniques adopted in the previous paragraph, we can
analytically locate the $P$-$AQ$ continuous transition line reported in 
Figs.~\ref{pdkm3} and \ref{pdkm5}:
\begin{equation}
\frac{1}{2} \beta^2 J^2 + \beta K =  \frac{1}{d(d-1)}   \virgula
\end{equation}
where $d$ satisfies the equation:
\begin{equation}
d \exp \left(\frac{1}{1-d} \right) = (1-d) \exp(\beta \mu) \ponto
\end{equation} 
Analogously, the $AQ$-$AG$ continuous transition line is given by: 
\begin{equation}
\frac{T^2}{J^2}=d_A d_B  \virgula
\end{equation}
where $d_A$ and $d_B$ satisfy the saddle point equations with $q_A=q_B=0$.

Finally, the $AG$-$SG$  transition line can be found expanding the
saddle point equations in the small quantities  $q_A-q_B$, $d_A-d_B$,
representing the staggered order parameter.

The stability of the solutions presented here can be studied  
by evaluating the eigenvalues of the matrix of gaussian fluctuations. 
The eigenvalues structure
is analogous to the one found in the SK  model  \cite{SK,AT} with
non zero magnetic field since here the chemical potential plays a 
similar role \cite{Arenzon,Schreiber,daCosta}. 
However, in this case the eigenvalues may be negative or even 
complex \cite{LagedeAlmeida,Salinas94,daCosta}.
The numerical study of their behavior \cite{daCosta} 
shows that the paramagnetic and antiquadrupolar phases are inside the 
stability region while both glass phases ($AG$ and $SG$) are 
unstable.
The critical line $P$-$AQ$, $P$-$SG$ and $AQ$-$AG$ are found to be in 
the stability region, their location depending on the chemical potential, 
and should not change when replica symmetry is broken; 
while the transition line $AG$-$SG$ is found to 
be completely inside the instability region and its precise location,
beyond the aims of this paper,
may be found only within the Parisi scheme of RSB.

\section{Conclusions}

We have studied the global phase diagram of 
a spin glass version of the Blume-Emery-Griffiths
model where the bilinear couplings are quenched gaussian random
variables.
The particles, besides the steric effects due to complex molecular 
structure (represented by the possible spin orientations), are subject 
to a potential that may be either attractive or repulsive. 
These steric effects that prevent close packing configuration are 
common in materials like glasses and poured sand.

This model displays a large variety of interesting
critical behaviors. 
When the particles interaction is attractive 
or weakly repulsive the transition between the paramagnetic
and spin glass phase may be either continuous or 
discontinuous, depending on the value of the chemical 
potential; these different behaviors are separated by a 
tricritical line. For strong repulsive interaction between particles,
new different phases with a sub-lattice structure emerge:
the antiquadrupolar and the antiquadrupolar glassy phase.
These phases seems to be separated by a continuous transition 
whatever is the value of the chemical potential (at least in the
range of $K$ studied here).
When the quenched disorder is not symmetrically distributed, 
also a ferromagnetic phase or a ferrimagnetic phase 
may appear in the phase diagram.

A further  transition line where a dynamical instability
appears may also be found as happens in other spin glass 
models with discontinuous transition.
Indeed, many interesting properties emerge if we consider the {\it dynamical}
behavior of systems whose equilibrium properties are described by 
Hamiltonian (\ref{H}). For example, the model with  diffusive particle 
dynamics shows strong glassy behavior characterized by diverging 
relaxation times, vanishing 
diffusivity and breakdown of the Debye-Stokes-Einstein law \cite{NC}; 
while the spin-1 version of the model with Monte Carlo dynamics displays 
different aging regimes due to the existence of spin-glass 
transitions of different nature. 
In the region where the spin-glass transition is discontinuous the aging 
behavior is stable against non-relaxational perturbation \cite{CKPS} 
and quite similar to the one observed in non-relaxational dynamics of 
the spherical $p$-spin glass model \cite{CKLP}.
Furthermore when a gravitational term is added to the Hamiltonian (\ref{H}),
strong links appears with granular media and their complex dynamical
behavior as logarithmic compaction \cite{NCH}.

\mbox{}
\vspace{1cm}
\mbox{}

\noindent
{\bf Acknowledgements:} We are very grateful to R.M.C. de
Almeida, A. Coniglio, L. Cugliandolo, J. Kurchan, N. Lemke, 
Th. Nieuwenhuizen, L. Peliti and S. Salinas
for fruitful discussions and comments. We also thank
F.A. da Costa for letting us know their results prior to
publication and to C.S. Yokoi for providing reference
\cite{Urahata}. JJA thanks the warm hospitality of the
Dipartimento di Scienze Fisiche at Napoli where this work
were started and to CNPq for partial support.

\appendix
\section{The 1RSB Solution}

In this appendix we present the solution of the Blume-Emery-Griffiths
spin-glass model with one-step  of replica symmetry breaking (1RSB).
The following relations are the starting point to get the phase diagram 
corrections and the dynamical
transition line.
Without loss of generality
we consider the case $J_0=0$
where the magnetization
is zero.
Following the Parisi scheme,  the $n$ replicas are
divided in $n/m$ blocks containing $m$ replicas. 
Different replicas in the same block have overlap $q_1$ while
those in different blocks have overlap $q_0$. Thus,
for the case of attractive particles interaction ($K>0$),
the 1RSB free energy reads:
\begin{eqnarray}
\beta f_1 &=& -\frac{1}{4} \beta^2 J^2 \left[ (1-m) q_1^2 + m q_0^2 -d^2
\right] + \frac{1}{2} \beta K  d^2 - \ln 2 \nonumber\\
&&-\frac{1}{m} \int \Dzo \ln \int \Dzu (1+\e^{\Xi_1} \cosh \Omega_1)^m
\end{eqnarray}
where:
\begin{eqnarray}
\Omega_1&=&\beta J z_0 \sqrt{q_0} + z_1 \sqrt{q_1-q_0}  \\
\Xi_1 &=& \frac{\beta^2 J^2}{2} (d-q_1) +\beta (\mu+Kd)
\end{eqnarray}
The saddle point equations are:
\begin{eqnarray}
d   &=&  \int\Dzo \overline{ \left(\frac{\e^{\Xi_1} \cosh \Omega_1}{
1+ \e^{\Xi_1} \cosh \Omega_1}  \right)
} \label{d1} \\
q_0 &=& \int\Dzo \overline{ \left( \frac{\e^{\Xi_1} \sinh \Omega_1}{
1+ \e^{\Xi_1} \cosh \Omega_1} \right)
}^2 \label{q0} \\
q_1 &=& \int\Dzo \overline{\left( \frac{\e^{\Xi_1} \sinh \Omega_1}{
1+ \e^{\Xi_1} \cosh \Omega_1} 
\right)^2} \label{q1}  
\end{eqnarray} 
($d\geq q_1\geq q_0$) and $m$ satisfies the equation:
\begin{eqnarray}
\frac{1}{4} m^2 \beta^2 J^2 (q^2_1 - q^2_0) 
 -m\int\Dzo \overline{\ln (1+\e^{\Xi_1} \cosh \Omega_1)}  
\nonumber\\
+ \int\Dzo\ln \int \Dzu (1+\e^{\Xi_1} \cosh \Omega_1)^m = 0
\label{m}
\end{eqnarray}
Here the overbar denotes the average:
\begin{equation}
\overline{X} = \frac{\int \Dzu (1+ \e^{\Xi_1} \cosh \Omega_1)^m X }{ 
\int \Dzu (1+ \e^{\Xi_1} \cosh \Omega_1)^m} 
\end{equation}
Following the same procedure, for the case of repulsive particles
interaction ($K<0$)
we obtain:
\begin{eqnarray}
\beta f_1 = &-&\frac{1}{4} \beta^2 J^2 \left[ (1- m) q_{1A}q_{1B}
+  m q_{0A} q_{0B} \right] \\   \nonumber
 &-& \frac{1}{4}  \left[ \beta^2 J^2 + 2\beta K \right]  d^2  - \ln 2  
 \nonumber \\ 
&-& \frac{1}{2m} \sum_{\alpha=A,B}
\int \Dzo \ln \int \Dzu A_{\alpha}^m(z_0,z_1)
\end{eqnarray}
where:
\begin{eqnarray*}
A_{\alpha}(z_0,z_1) &=& 1 +  
\e^{\Xi_{1\alpha}} \cosh (\beta J z_0 \sqrt{q_{0\alpha}} 
+ \beta J z_1 
\sqrt{q_{1\alpha}-q_{0\alpha}})  \\
\Xi_{1\alpha}     &=& \frac{\beta^2 J^2}{2} (d-q_{1\alpha}) 
+\beta (\mu+Kd) 
\end{eqnarray*}
It's  straightforward, but tedious exercise, to obtain the saddle point 
equations.

\break

\break

\break

\break

\break
    
\begin{center}
\begin{figure}[h]
\epsfig{file=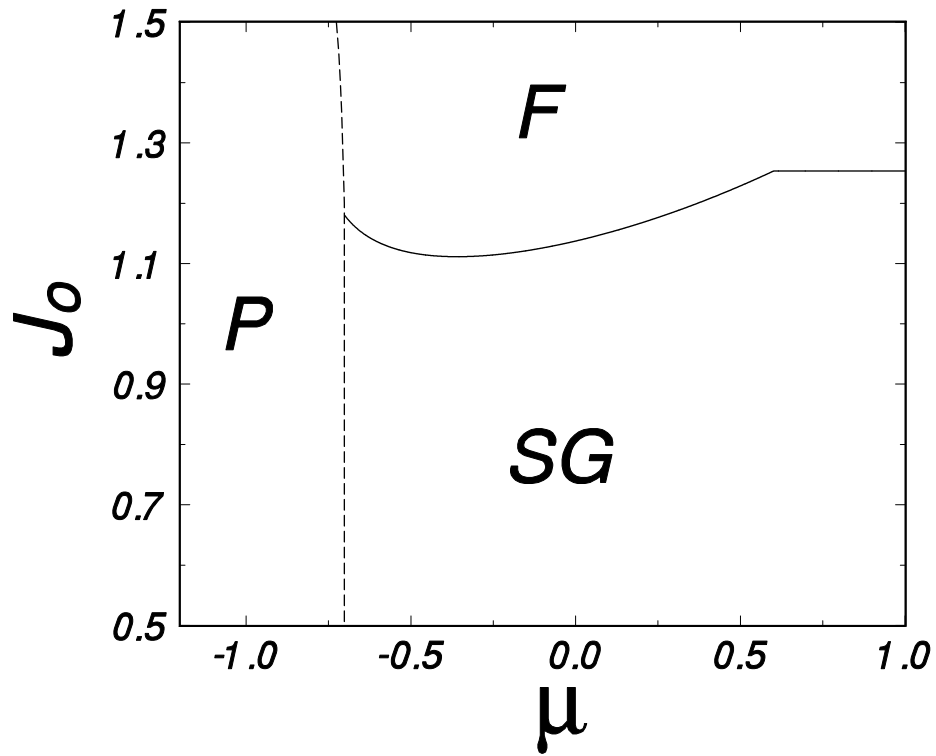,width=8.5cm,angle=270} 
\vspace{3.5cm}
\caption{Phase diagram at $T=0$ for $K=-1$ showing the 
critical {\it endpoint} when both curves meet.}
\end{figure}
\end{center}

\break

\break

\break

\break

\begin{thebibliography}{40}
    
\bibitem{BEG} M. Blume, V.J. Emery and R.B. Griffiths 1971 
        {\em Phys. Rev. A}   {\bf 4} 1071.

\bibitem{Tanaka} M. Tanaka and T. Kawabe 1985 {\em J. Phys.
        Soc. Japan} {\bf 54} 2194.

\bibitem{Salinas89} R. Os\'orio, M.J. de Oliveira and S.R. Salinas 1989
        {\em J. Phys. C}   {\bf 1} 6887.
    
\bibitem{HB} W. Hoston and A.N. Berker 1991 
        {\em Phys. Rev. Lett.} {\bf 67} 1027.

\bibitem{Branco} N.S. Branco 1996 {\em Physica A} {\bf 232} 477.

\bibitem{Akheyan} A.Z. Akheyan and N.S. Ananikian 1996
        {\em J. Phys. C} {\bf 29} 721.

\bibitem{MPV} M. M\'ezard, G. Parisi and M. Virasoro 1987 {\em Spin Glass 
        Theory and Beyond}, World Scientific (Singapore).
     
\bibitem{GS} S.K. Ghatak and D. Sherrington 1977 
        {\em J. Phys. C} {\bf 10} 3149.
    

\bibitem{LagedeAlmeida} E.J.S. Lage and J.R.L. de Almeida 1982 
        {\em J. Phys. C} {\bf 15} L1187.
  
\bibitem{MottishawSherrington} P. Mottishaw and D. Sherrington 1985
        {\em J. Phys. C} {\bf 18} 5201. 
    
\bibitem{Yokota} T. Yokota 1992 {\em J. Phys. C} {\bf 4} 2615.

\bibitem{Salinas94} F.A. da Costa, C.S.O. Yokoi and S.R. Salinas 1994
        {\em J. Phys. A}  {\bf 27} 3365.

\bibitem{Urahata} C.S. Yokoi and S.M. Urahata 1995 in {\em
        Proceedings of the Symposium on Statistical Mechanics
	  and Information Sciences}, University of Tohoku, Japan, 53.
            
\bibitem{Arenzon} J.J. Arenzon, M. Nicodemi and M. Sellitto 1996
        {\em J. Physique I}  {\bf 6} 1143. 

\bibitem{NC} M. Nicodemi and A. Coniglio 1997 {\em J. Phys. A}
        in press.

\bibitem{Coniglio} A. Coniglio 1993 {\em J. Phys. IV, Colloq. C1}, 
        {\bf 3} 1; 1994 {\em Il Nuovo Cimento D} {\bf 16} 1027.
    
\bibitem{Kirk} T.R. Kirkpatrick and D. Thirumalai 1987 
        {\em Phys. Rev. B} {\bf 36} 5388.

\bibitem{softpotts} D. Thirumalai and T.R. Kirkpatrick 1988 {\em Phys.
      Rev. B} {\bf 37} 5342 and {\em Phys. Rev. B} {\bf 38} 4881.
	
\bibitem{Felix} E. De Santis, G. Parisi and F. Ritort
	1995   {\em J. Phys. A} {\bf 28} 3025.
	
\bibitem{pspin} A. Crisanti and H.-J. Sommers 1992 {\em Z. Phys.  B}
      {\bf 87} 341; A. Crisanti, H. Horner and H.-J. Sommers 1993
	{\em Z. Phys. B} {\bf 92} 257.

\bibitem{thankstoTheo} We thank Theo Nieuwenhuizen for 
        pointing this out to us.

\bibitem{TAP} D.J. Thouless, P.W. Anderson and R.G. Palmer 1977 
		{\em Phil. Mag.} {\bf 35}   593.
 
\bibitem{Maurothesis} M. Sellitto, in preparation.

\bibitem{Heiko} A.J. Bray and M.A. Moore 1980 {\em J. Phys. C}
		{\bf 13} L469;
		H. Rieger 1992 {\em Phys. Rev. B} {\bf 46} 14655.
	
\bibitem{KD} J. Kurchan, G. Parisi and M.A. Virasoro 1993 
	{\em J. Physique I} {\bf 3} 1819; 
	A. Crisanti and H.-J. Sommers 1995  {\em J. Physique I} {\bf 5} 805;
	 T.R.~Kirkpatrick and D. Thirumalai 1995 {\em J. Physique I} 
        {\bf 5} 777.

\bibitem{marginal} 
	H.J. Sommers 1983 {\em Z. Phys. B} {\bf 50} 97;
 	I. Kondor and C. De Dominicis 1986 {\em Europhys. Lett.}
	{\bf 2} 617;
	M. Freixa-Pascual and H. Horner 1990 {\em Z. Phys. B} 
	{\bf 80} 95.


\bibitem{CuKu} L.F. Cugliandolo and J. Kurchan 1993 {\em Phys. Rev. Lett.} 
        {\bf 71} 173; 1994 {\em J. Phys. A } {\bf 27} 5749; 
        1995 {\em Phil. Mag. B} {\bf 71} 501.
	
\bibitem{FraMe} S. Franz and M. Mezard 1994 {\em Europhys. Lett.} 
	{\bf 26}   209.

\bibitem{CuDou} L.F. Cugliandolo and P. Le Doussal 1996 {\em Phys. Rev. E}
        {\bf  53} 1525.
	  
\bibitem{SK} D. Sherrington and S. Kirkpatrick 1975 
        {\em Phys. Rev. Lett.} {\bf 35} 1792.

\bibitem{AT} J.R.L. de Almeida and D.J. Thouless 1978 
        {\em J. Phys. A} {\bf 11} 983.

\bibitem{Schreiber} G.R. Schreiber, cond-mat/9612189.

\bibitem{daCosta} F.A. da Costa, F.D. Nobre and C.S.O. Yokoi 1997
        {\em J. Phys. A}, in press.

\bibitem{FKS} I.Y. Korenblit  and E.F. Shender 1986 {\em Sov. Phys. JETP}
        {\bf 62} 1030; Y.V. Fyodorov, I.Y. Korenblit and E.F. Shender 1987
        {\em J. Phys. C} {\bf 20} 1835.

\bibitem{CKPS} L. Cugliandolo, J. Kurchan, L. Peliti and M. Sellitto,
        In preparation.
    
\bibitem{CKLP} L. Cugliandolo, J. Kurchan, P. Le Doussal and L. Peliti, 
        cond-mat/9606060, {\em Phys. Rev. Lett.} in press.

\bibitem{NCH} M. Nicodemi, A. Coniglio and H. J. Herrmann, in
        cond-mat/9606097 and to be published.

\end{thebibliography}
\end{document}